\documentclass{article}
\usepackage{ijcai19}

\usepackage{times}
\usepackage{soul}
\usepackage{url}
\usepackage[utf8]{inputenc}
\usepackage[small]{caption}
\usepackage{graphicx}
\usepackage{amsmath}
\usepackage{booktabs}
\usepackage{algorithm}
\usepackage{times}  
\usepackage{algpseudocode}
\usepackage{subcaption}
\usepackage{color}
\usepackage{amssymb}  
\usepackage{mathtools}
\usepackage{xcolor}
\usepackage{cleveref}
\usepackage{enumitem}

\title{Universal Policies to Learn Them All}

\author{ \parbox{3 in}{\centering Hassam Ullah Sheikh\\
        Department of Computer Science\\
        University of Central Florida\\
        Orlando, Florida, United States\\
        {\tt\small hassam.sheikh@knights.ucf.edu}}
        \hspace*{ 0.5 in}
        \parbox{3 in}{\centering Ladislau B{\"o}l{\"o}ni\\
        Department of Computer Science\\
        University of Central Florida\\
        Orlando, Florida, United States\\
        {\tt\small lboloni@cs.ucf.edu}}
}

\begin{document}

\maketitle

\begin{abstract}
  We explore a collaborative and cooperative multi-agent reinforcement learning setting where a team of reinforcement learning agents attempt to solve a single cooperative task in a multi-scenario setting.  We propose a novel multi-agent reinforcement learning algorithm inspired by universal value function approximators that not only generalizes over state space but also over a set of different scenarios. Additionally, to prove our claim, we are introducing a challenging 2D multi-agent urban security environment where the learning agents are trying to protect a person from nearby bystanders in a variety of scenarios. Our study shows that state-of-the-art multi-agent reinforcement learning algorithms fail to generalize a single task over multiple scenarios while our proposed solution works equally well as scenario-dependent policies. 
\end{abstract}

\section{Introduction}
\label{sec:Introduction}

Recent research in deep reinforcement learning (RL) has led to wide range of accomplishments in learning optimal policies for sequential decision making problems. These accomplishments include training agents in simulated environments such as playing Atari games~\cite{Mnih-2015-Nature}, beating the best players in board games like Go and Chess~\cite{Silver-2016-Nature} as well as learning to solve real world problems. Similar to single agent reinforcement learning, multi-agent reinforcement (MARL) is also producing break through results in challenging collaborative-competitive environments such as~\cite{OpenAI-2018,Max-2018-ARXIV,Liu-2019-ICLR}. 

The success in reinforcement learning has prompted interest in more complex challenges as well as a shift towards cases in which an agent tries to learn multiple tasks in a single environment. Formally this paradigm of learning is known as multitask reinforcement learning~\cite{Teh-2017-NIPS}. The essence of multitask reinforcement learning is to simultaneously learn multiple tasks jointly to speed up learning and induce better generalization by exploiting the common structures among multiple tasks.  

Despite having success in single agent multitask reinforcement learning~\cite{Borsa-2019-ICLR,Teh-2017-NIPS}, multitask multi-agent reinforcement learning has been explored in only one recent study~\cite{Omidshafiei-2017-ICML}. In this paper, we are exploring an opposite problem where multiple reinforcement learning agents are trying to master a single task across multiple scenarios. Consider multiple RL agents trying to master a single task in multiple scenarios. In order for the agents to generalise, they need to able to identify and exploit common structure of the single task under multiple scenarios. One possible structure is the similarity between the solutions of the single task over multiple scenarios either in the policy space or associated value-function space. For this, we build our solution upon two frameworks. The first framework is {\em universal value function approximators} (UVFAs) by~\cite{Schaul-2015-ICML} and the second framework is {\em multi-agent deep deterministic policy gradient} (MADDPG) by~\cite{Lowe-2017-NIPS}. UVFAs are extension of value functions that also include the notion of a task or a scenario thus exploiting common structure in associated optimal value functions. MADDPG is a multi-agent reinforcement learning algorithm that uses the centralized training and distributed testing paradigm to stabilize learning.  The outcome of combining these two frameworks is a solution for multi-agents that learn to generalize over both state space and a set of multiple scenarios.

To investigate the emergence of collaboration in multi-scenario learning for multi-agent learning agents, we designed a challenging environment with simulated physics in Multi-Agent Particle Environment~\cite{Mordatch-2017-ARXIV}. We have developed 4 different simulated scenarios representing a challenging urban security problem of providing physical protection to VIP from nearby bystanders of more than one different class.  The complexity in our environment arises primarily from the different moving patterns of these bystanders that a standard state-of-the-art MARL algorithm such as MADDPG~\cite{Lowe-2017-NIPS} fail to capture.
The goal here is to learn a stable and a consistent multi-agent cooperative behavior across all the known scenarios. Here, we are not dealing with unknown scenarios. 

\section{Background}
\label{sec:Background}
Partially observable Markov Game~\cite{Littman-1994-ICML} is a multi-agent extension of MDP characterized by $\mathcal{S}$, $N$ agents with partial observations $\mathcal{O}=\left \{ \mathcal{O}_1, \ldots, \mathcal{O}_N\right  \}$ of the environment with a collective action space of $\mathcal{A}=\left \{ \mathcal{A}_1, \ldots, \mathcal{A}_N\right  \}$, a reward function $\mathcal{R}$ and a state transition function $\mathcal{T}$. At every time step, each agent chooses an action $a_i$ from it's policy $\pi$ parameterized by $\theta_i$ conditioned on its private observation $o_i$ i.e $a_i= \pi_{\theta_{i}}\left(o_i\right)$ and receives a reward $r_i=\mathcal{R}\left(s, a_i\right)$ 
The goal of each agent is to maximise its own total expected return $\mathbb{E}\left[G_i\right]=\mathbb{E}\left[\sum_{t=0}^{T}\gamma^{t}r_i^t \right]$ where $r_i^t$ is the collected reward by agent $i$ at time $t$.

\subsection{Policy Gradients}

Policy gradient methods have been shown to learn the optimal policy in a variety of reinforcement learning tasks. The main idea behind policy gradient methods is to maximize the objective function by parameterizing the policy $\pi$ with $\theta$ and updating the policy parameters in the direction of the gradient $\nabla J{\left(\theta\right)}$  of the objective function $J\left(\theta\right)=\mathbb{E}\left[{G}^{t}\right]$. The gradient is defined as
\[
\nabla J{\left(\theta\right)}=\mathbb{E}\left[\nabla_{\theta}\log\pi_{\theta}\left(a|s\right)Q^{\pi}\left(s,a\right)\right]
\]

\cite{Silver-2014-ICML} has shown that it is possible to extend the policy gradient framework to deterministic policies {\em i.e.} $\pi_{\theta}:\mathcal{S}\rightarrow\mathcal{A}$.
In particular we can write $\nabla J\left(\theta\right)$ as
\[
\nabla J{\left(\theta\right)}=\mathbb{E}\left[\nabla_{\theta}\pi\left(a|s\right)\nabla_{a}Q^{\pi}\left(s,a\right)|_{a=\pi\left(s\right)}\right]
\]
A variation of this model, Deep Deterministic Policy Gradients (DDPG)~\cite{Lillicrap-2015-ICLR} is an off-policy algorithm that approximates the policy $\pi$ and the critic $Q^{\pi}$  with deep neural networks. DDPG also uses an experience replay buffer alongside a target network to stabilize the training.

Multi-agent deep deterministic policy gradients (MADDPG)~\cite{Lowe-2017-NIPS} extends DDPG for the multi-agent setting where each agent has it's own policy. The gradient of each policy is written as
\[
\nabla J{\left(\theta_{i}\right)} = \mathbb{E}\left[\nabla_{\theta_{i}}\pi_{i}\left(a_{i}|o_{i}\right)\nabla _{a_i}Q_{i}^{\pi}\left(s,a_{1},\ldots,a_{N}\right)|_{a_i=\pi_i\left(o_i\right)}\right]
\]
\noindent where $s=\left(o_1, \ldots, o_N\right)$  and   $Q_{i}^{\pi}\left(s,a_{1},\ldots,a_{N}\right)$ is a centralized action-value function that takes the actions of all the agents in
addition to the state of the environment to estimate the Q-value for agent
$i$. Since every agent has it's own Q-function, the model allows the agents
to have different action spaces and reward functions. The primary motivation
behind MADDPG is that knowing all the actions of other agents makes
the environment stationary that helps in the stabilization of the training,  even though the policies of the agents change.

The Universal Value Function Approximator~\cite{Schaul-2015-ICML} is an extension of DQN~\cite{Mnih-2015-Nature} where it generalizes not only over a set of states but also on a set of goals. At the beginning of the episode, a state-goal pair is sampled from a probability distribution, the goal remains constant throughout the episode. At each timestep, the agent receives a pair of current state and goal and gets the reward $r^t = r_g\left(s^t, a^t\right)$. As a result, the Q-function is not only dependent on state-action pair but also on the goal. The extension of this approach is straight forward for DDPG~\cite{Marcin-2017-NIPS} and MADDPG.

\section{Multi-Agent Universal Policy Gradient}
We propose {\em multi-agent universal policy gradient}: a multi-agent deep reinforcement learning algorithm that learns distributed policies not only over state space but also over a set of scenarios.  

While generalization across multi-task in multi-agent reinforcement learning has been studied in~\cite{Omidshafiei-2017-ICML}, to our best knowledge we are the first one to consider the multi-scenario multi-agent deep RL system.
Our approach uses Universal Value Function Approximators~\cite{Schaul-2015-ICML} to train policies and value functions that take a state-scenario pair as an input. The outcome are universal multi-agent policies that are able to perform on multiple scenarios as well as policies that are trained separately.

The main idea is to represent the different value function approximators for each agent $i$ by a single unified value function approximator that generalizes of over both state space and a set of scenarios. For agent $i$ we consider $V_i\left(s, g; \phi\right)\approx V_{i_g}^*\left(s\right)$ or
$Q_i\left(s, a, g; \phi\right)\approx Q_{i_g}^*\left(s, a\right)$ that approximate the optimal unified value functions over multiple scenarios and a large state space. These value functions can be used to extract policies implicitly or as critics for policy gradient methods.

The learning paradigm we used is similar to the centralized training with decentralized execution during testing used by~\cite{Lowe-2017-NIPS}. In this setting, additional information is provided for the agents during training that is not available during test time. Thus, extracting policies from value functions is not feasible in this model. However, the value functions can be used as critics in a multi-agent deep deterministic policy gradient setting.


Concretely, consider an environment with $N$ agents with policies $\pmb{\pi}=\{\pmb{\pi_1}, \ldots, \pmb{\pi_N}\}$ parameterized by $\pmb{\theta}=\{\pmb{\theta_1}, \ldots, \pmb{\theta_N}\}$ then the \textit{multi-agent deep deterministic policy gradient} for agent $i$ can written as
\[
\nabla J{\left(\theta_{i}\right)}=\mathbb{E}\left[\nabla_{\theta_{i}}\pi_{i}\left(a_{i}|o_{i}\right)\nabla _{a_i}Q_{i}^{\pi}\left(s,a_{1},\ldots,a_{N}\right)|_{a_i=\pi_i\left(o_i\right)}\right]
\]
\noindent where $s=\left(o_1, \ldots, o_N\right)$  and   $Q_{i}^{\pi}\left(s,a_{1},\ldots,a_{N}\right)$ is a centralized action-value function parameterized by $\phi_i$ that takes the actions of all the agents in addition to the state of the environment to estimate the Q-value for agent $i$. We extend the idea of MADDPG with universal functional approximator, specifically we augment the centralized critic with an embedding of the scenario. Now the modified policy gradient for each agent $i$ can be written as
\begin{align}
\label{eq:maupg}
\begin{split}
 \nabla J{\left(\theta_{i}\right)} &= \mathbb{E}_{s,a, g\thicksim \mathcal{D}}\Bigg[\nabla_{\theta_{i}}\pi_{i}\left(a_i|o_{i}, g \right) \\
   								 &\qquad  \nabla_{a_i}Q_{i}^{\pi}\left(s,a_{1},\ldots,a_{N}, g\right)  \Bigg]
\end{split}
\end{align}
\noindent where ${a_i=\pi_i\left(o_i, g\right)}$ is action from agent $i$ following policy $\pi_i$ and $\mathcal{D}$ is the experience replay buffer. The centralized critic is $Q_{i}^{\pi}$ is updated as:
\begin{equation*}
\mathcal{L}\left(\phi_i\right)=\mathbb{E}_{s,a,r, s',g}\left[\left(  Q_{i}^{\pi}\left(s,a_{1},\ldots,a_{N}, g\right)\right) -y \right]
\end{equation*}
\noindent where $y$ is defined as:
\begin{equation*}
y=r_i^g+\gamma Q_{\phi'_i}^{\pi'}\left(s',a'_{1},\ldots,a'_{N}, g\right)|_{a'_i=\pi'_i\left(o_i', g\right)}
\end{equation*}
\noindent where $\pmb{\pi'}=\{\pmb{\pi'_1}, \ldots, \pmb{\pi'_N}\}$ are target policies parameterized by $\pmb{\theta'}=\{\pmb{\theta_1'}, \ldots, \pmb{\theta'_N}\}$.

\begin{figure}
    \centering
    \begin{subfigure}[t]{0.6\columnwidth}
	\label{subfig: policy}
        \centering
        \includegraphics[height=1.5in]{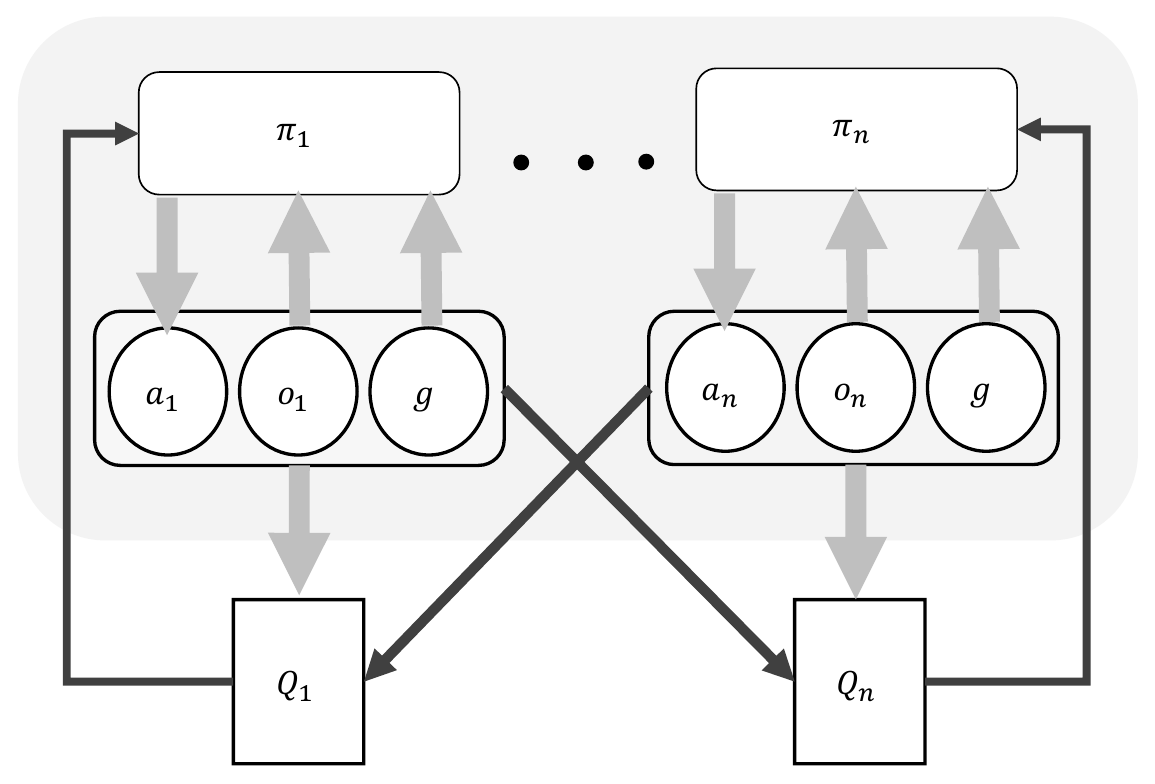}
        \caption{An overview of the mutli-agent decentralized actor with centralized critic represented by a universal value function approximator.}
    \end{subfigure}%
    ~
    \begin{subfigure}[t]{0.35\columnwidth}
	\label{subfig: critic}
        \centering
        \includegraphics[height=1.5in]{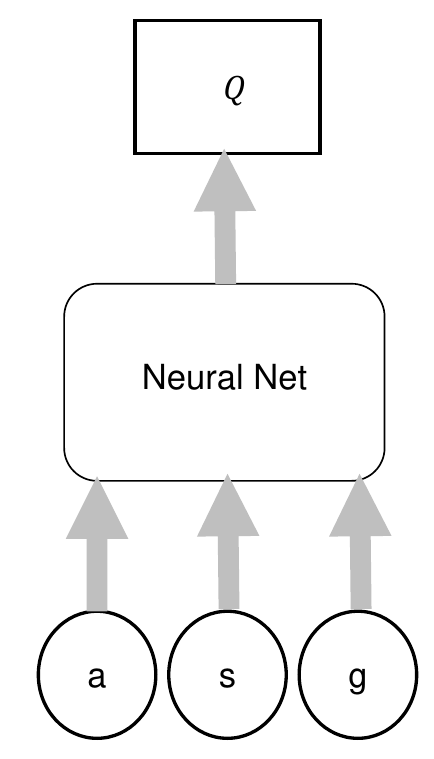}
        \caption{Representation of the single centralized critic by a universal function approximator.}
    \end{subfigure}%
    ~
    \caption{An overview of the multi-agent universal policy gradient architecture where both actors and critics are augmented with the goal that the agents are trying to achieve.}
\label{fig:maupg}
\end{figure}

The overall algorithm to which we refer as \textit{multi-agent universal policy gradient (MAUPG)} is described in~\cref{alg:1}. Additionally, we refer the learnt policies as universal policies. The overview of the architecture can be seen in ~\cref{fig:maupg}.

\begin{algorithm}[H]
  \caption{Multi-agent Universal Policy Gradient}
  \begin{algorithmic}[1]
  \State Sample a random scenario $g$
  	\For{episode = 1 to $M$}
        \For{t = 1 to episode--length}
   			\State For each agent $i$, select action $a_{i}^{t}=\pi_{\theta_{i}}\left(o^{t}_{i}, g\right)$
            \State Execute actions $\textbf{a}_{t}=\left[a^{t}_{1},\ldots,a^{t}_{N}\right]$
            \State For each agent $i$, get next observation $o_{i}^{t+1}$
        \EndFor
        \State Sample an additional scenario $k$
        \For{t = 1 to episode--length}
	 \For{agent i = 1 to N}
	    \State Get reward $r_i^t  \coloneqq r^{g}_{i}\left(o^{t}_{i},a^{t}_{i}\right)$
        \State Store $\left(o^{t}_{i}, a^{t}_i, r_{i}^{t}, o^{t+1}_{i}, g\right)$ in  replay buffer
        \State /* Hindsight Replay */
        \State Get reward $r_i^t  \coloneqq r^{k}_{i}\left(o^{t}_{i},a^{t}_{i}\right)$ \label{alg:hr_reward}
	    \State Store $\left(o^{t}_{i}, a^{t}_i, r_{i}^{t}, o^{t+1}_{i}, k\right)$ in  replay buffer \label{alg:hr}
        \EndFor
        \EndFor
        \State Set $g = k$
        \For{agent $i = 1$ to $N$}
        \State Sample minibatch of size S $\left(\textbf{s}^{j}, \textbf{a}^{j}, \textbf{r}^{j}, \textbf{s}^{'j}, \textbf{g}^{j}\right)$
        \State $
        \textbf{w}\coloneqq\left(\pi'_{\theta_{i}}(o'^{j}_{i}, g^j),\ldots,\pi'_{\theta_{N}}(o'^{j}_{N}, g^j) \right)$
       \State $
        \left(a'_{1},\ldots,a'_{N} \right) \coloneqq  \textbf{w}$
        \State Set $y^j = r_i^j+\gamma Q_{\phi'_i}^{\pi'}\left(\textbf{s}^{'j},a'_{1},\ldots,a'_{N}, \textbf{g}^{j}\right)$
        \State Update critic by minimizing  \[ \frac{1}{S} \sum_j \left(y^j -  Q_{\phi_i}^{\pi}\left(\textbf{s}^{j},a^{j}_{1},\ldots,a^j_{N}, \textbf{g}^{j}\right) \right)\]
         \State  $\theta_i +=  \sum_j \frac{  \nabla_{\theta_{i}}\pi_{i}\left(a_{i}|o_{i}^j, g^{j}\right)\nabla _{a_i}Q_{\phi_i}^{\pi}\left(\textbf{s}^{j},a^{j}_{1},..,a^j_{N}, \textbf{g}^{j}\right)}{S}$

        \EndFor
        \State Update target network parameters for each agent $i$
        \begin{align*}
        	\theta'_i \leftarrow \tau\theta_i + \left(1-\tau\right)\theta'_i\\
        \phi'_i \leftarrow \tau\phi_i + \left(1-\tau\right)\phi'_i
        \end{align*}

    \EndFor
  \end{algorithmic}
  \label{alg:1}
\end{algorithm}

\begin{figure*}
    \centering
    \begin{subfigure}[t]{0.24\textwidth}
        \centering
        \fbox{\includegraphics[height=1.55in]{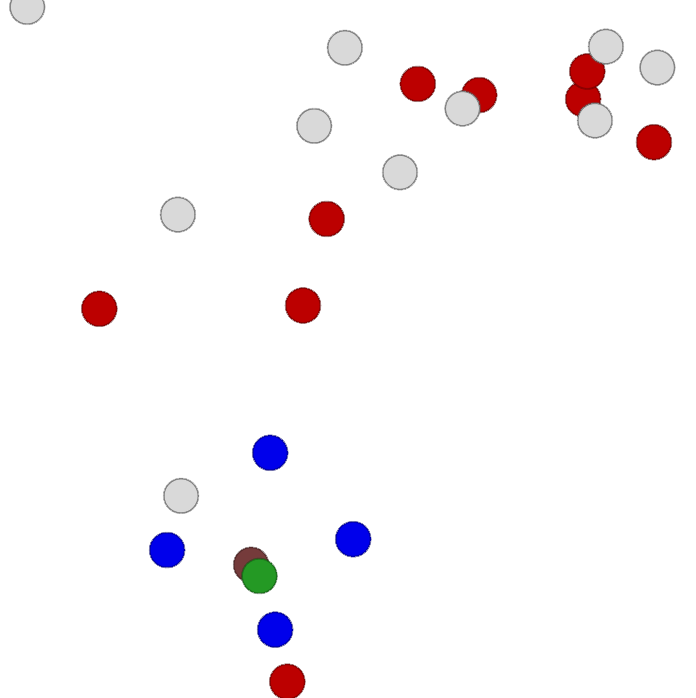}}
        \caption{Random Landmarks \label{subfig:rl}}
    \end{subfigure}%
    ~
    \begin{subfigure}[t]{0.24\textwidth}
        \centering
        \fbox{\includegraphics[height=1.55in]{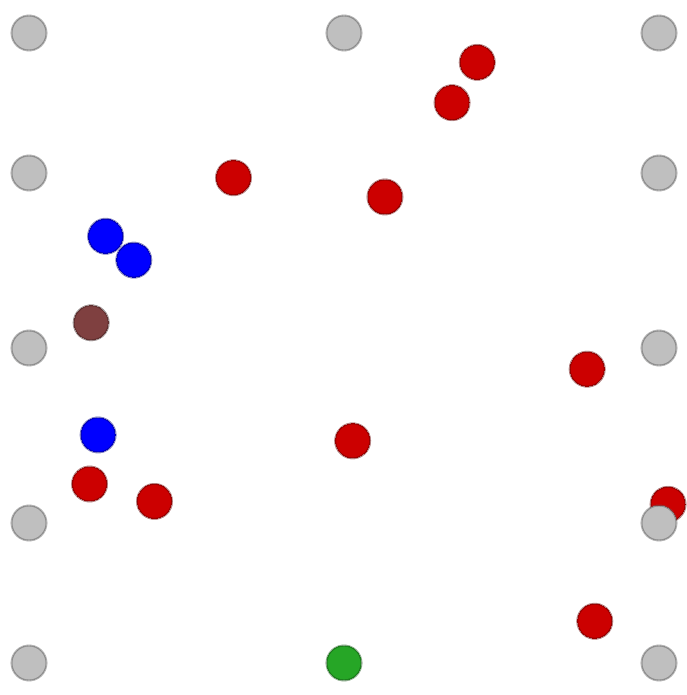}}
        \caption{Shopping Mall \label{subfig:sm}}
    \end{subfigure}%
    ~
    \begin{subfigure}[t]{0.24\textwidth}
        \centering
        \fbox{\includegraphics[height=1.55in]{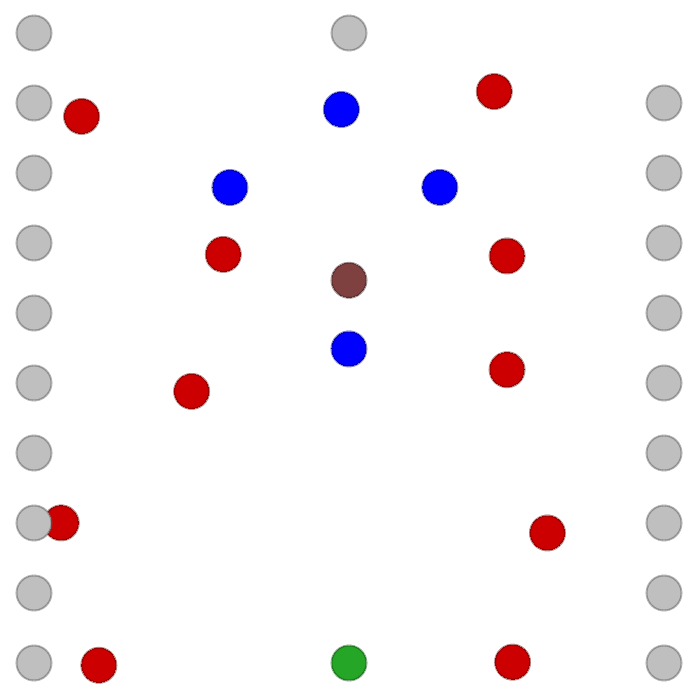}}
        \caption{Street \label{subfig:st}}
    \end{subfigure}%
    ~
    \begin{subfigure}[t]{0.24\textwidth}
        \centering
        \fbox{\includegraphics[height=1.55in]{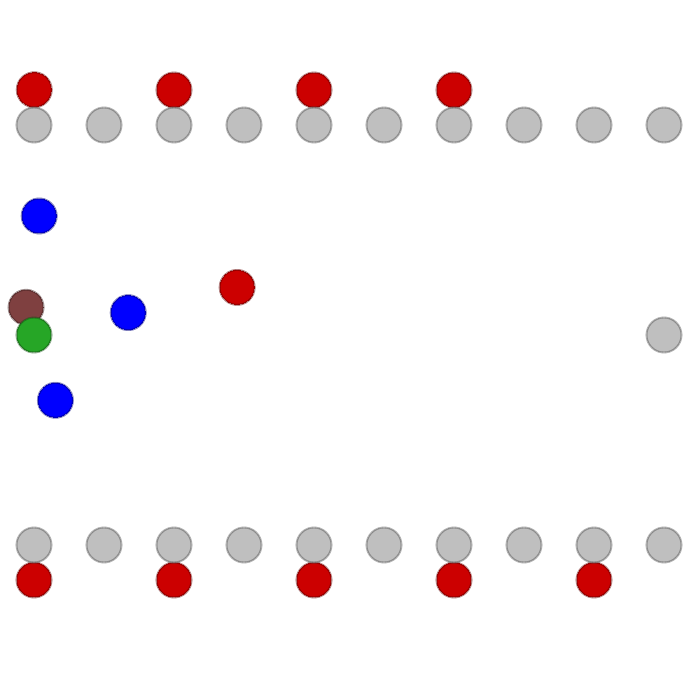}}
        \caption{Pie-in-the-face \label{subfig:pf}}
    \end{subfigure}
    \caption{Visual representation of the four different scenarios. Emergence of complex behavior can be clearly seen  where the bodyguards(in blue) have positioned themselves between the VIP(in brown) and the bystanders(in red) shielding from potential threat.}
    \label{fig:environments}
\end{figure*}

\begin{figure*}[h]
    \centering
    \begin{subfigure}[t]{0.25\textwidth}
        \centering
       \includegraphics[width=1.8in,height=1.6in]{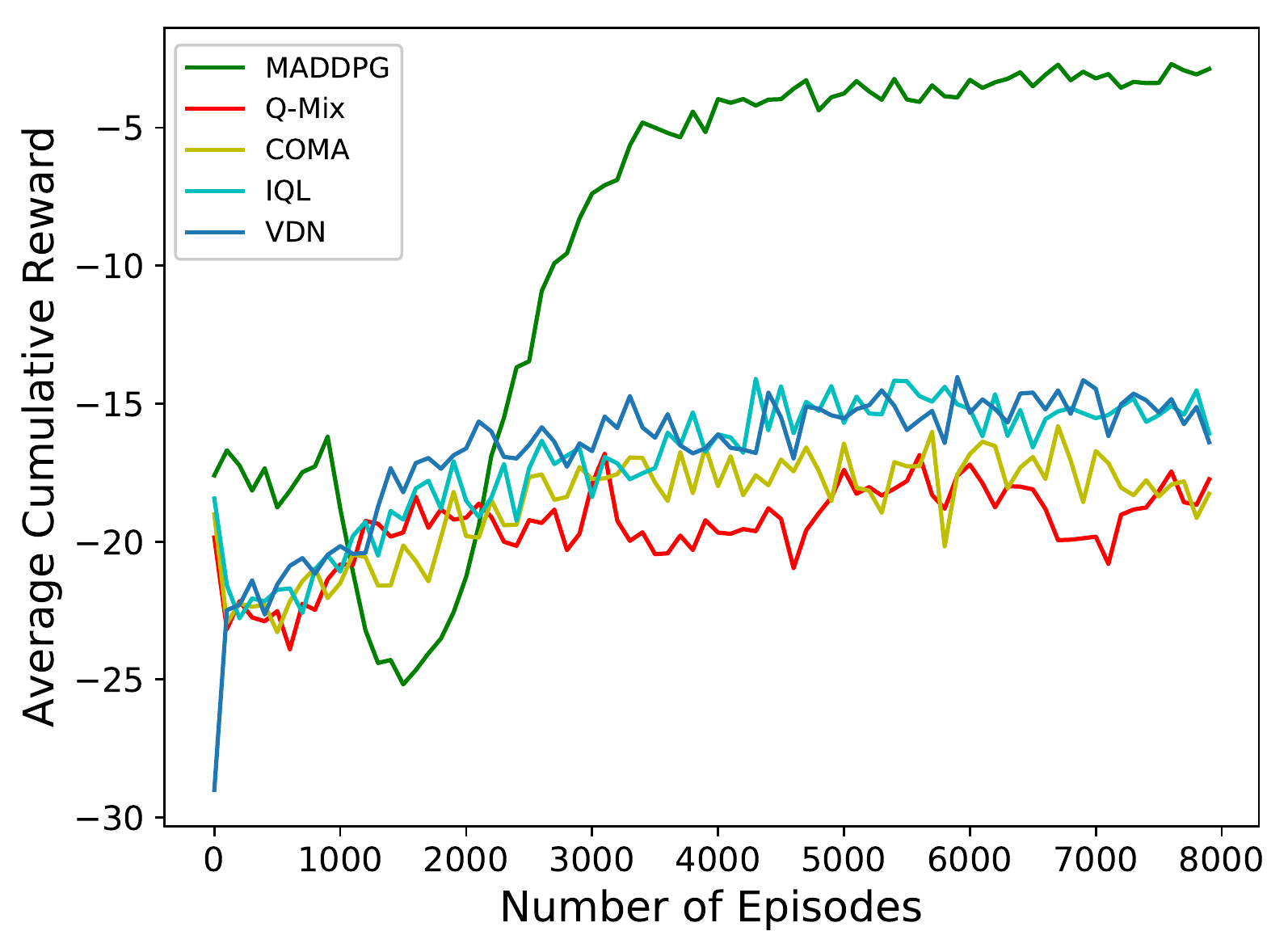}
        \caption{Random Landmarks \label{subfig:rl_reward}}
    \end{subfigure}%
    ~
    \begin{subfigure}[t]{0.25\textwidth}
        \centering
       \includegraphics[width=1.8in,height=1.6in]{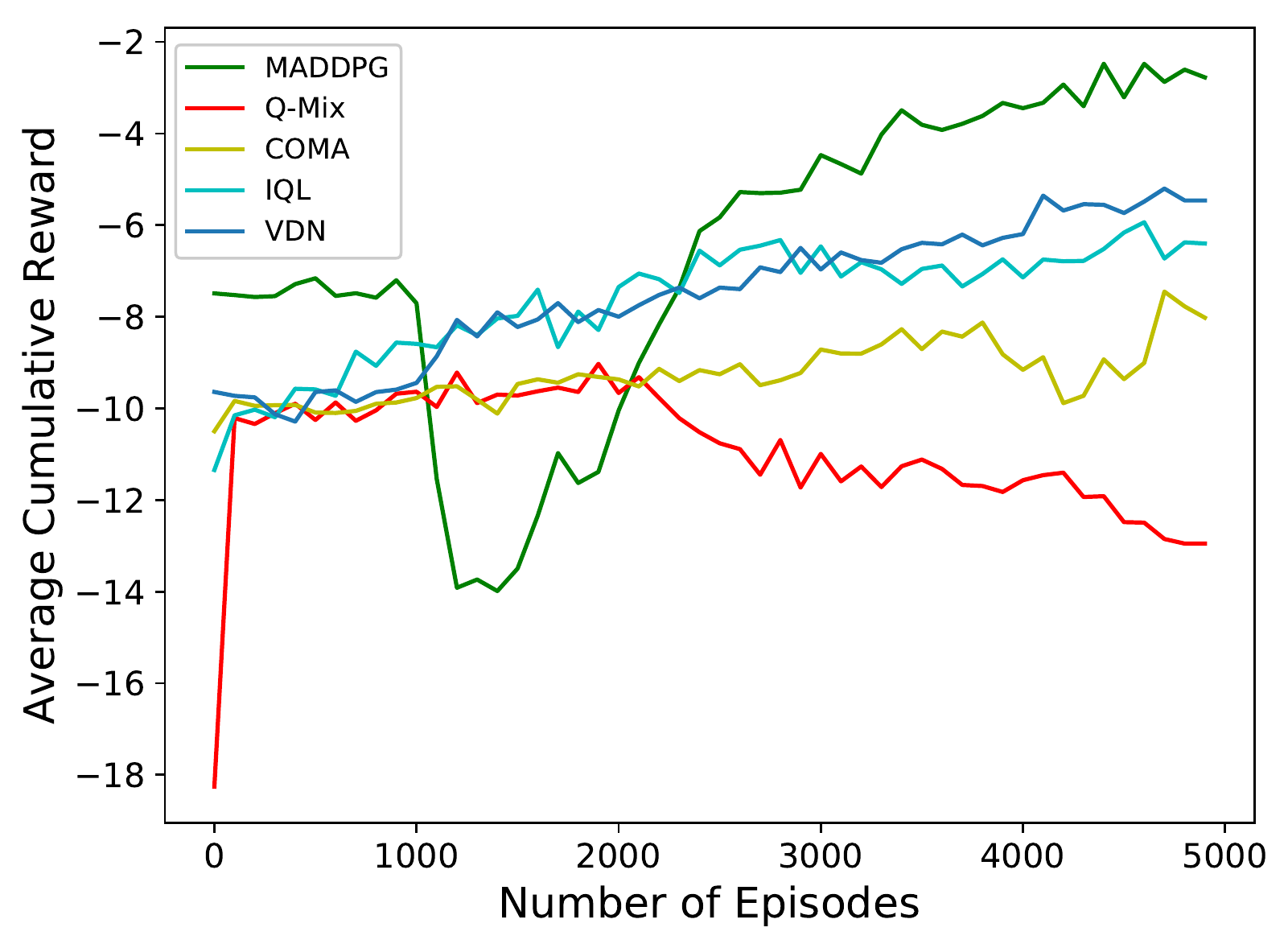}
        \caption{Shopping Mall \label{subfig:sp_reward}}
    \end{subfigure}%
    ~
    \begin{subfigure}[t]{0.25\textwidth}
        \centering
       \includegraphics[width=1.8in,height=1.6in]{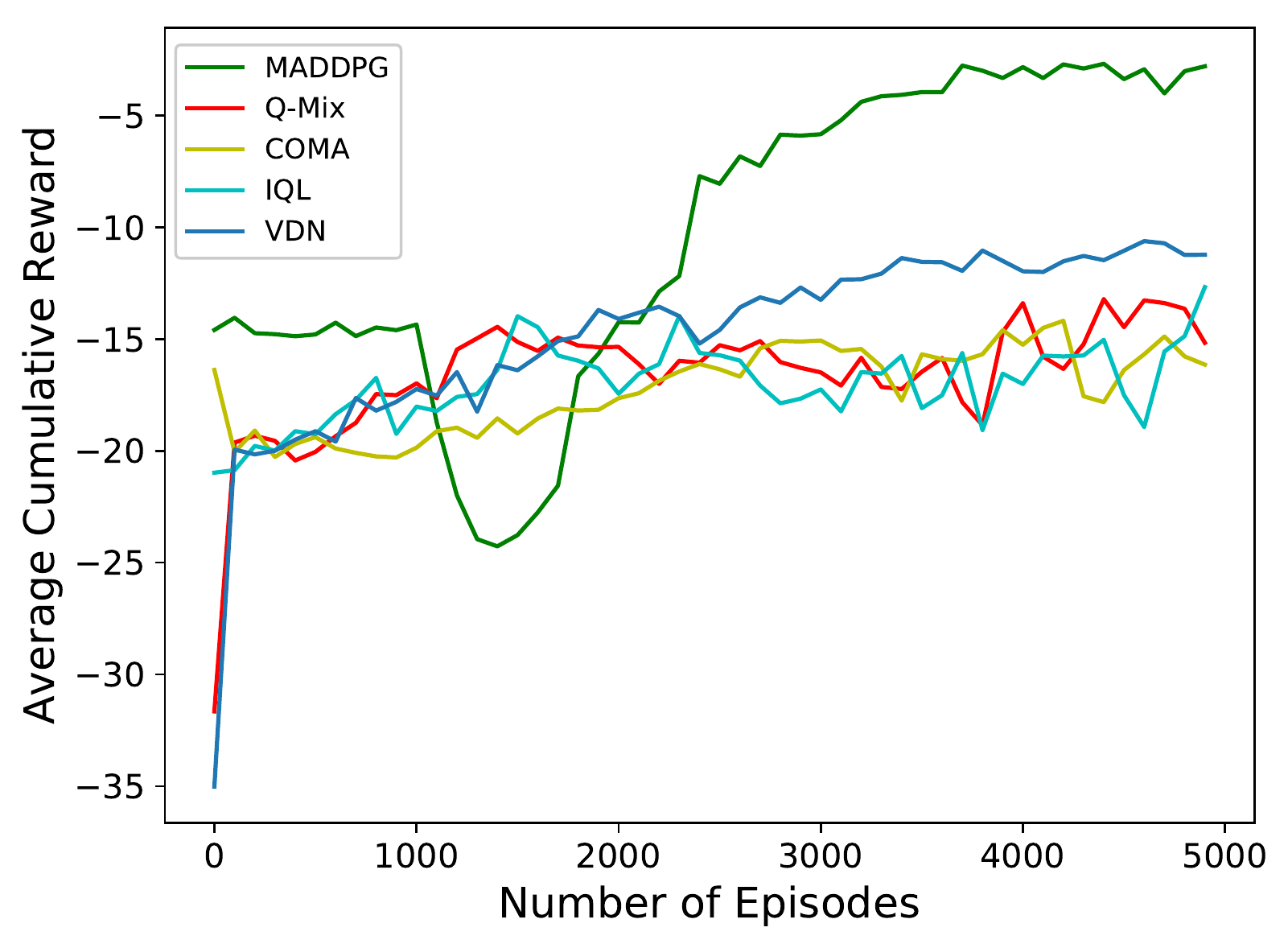}
        \caption{Street \label{subfig:st_reward}}
    \end{subfigure}%
    ~
    \begin{subfigure}[t]{0.25\textwidth}
        \centering
       \includegraphics[width=1.8in,height=1.6in]{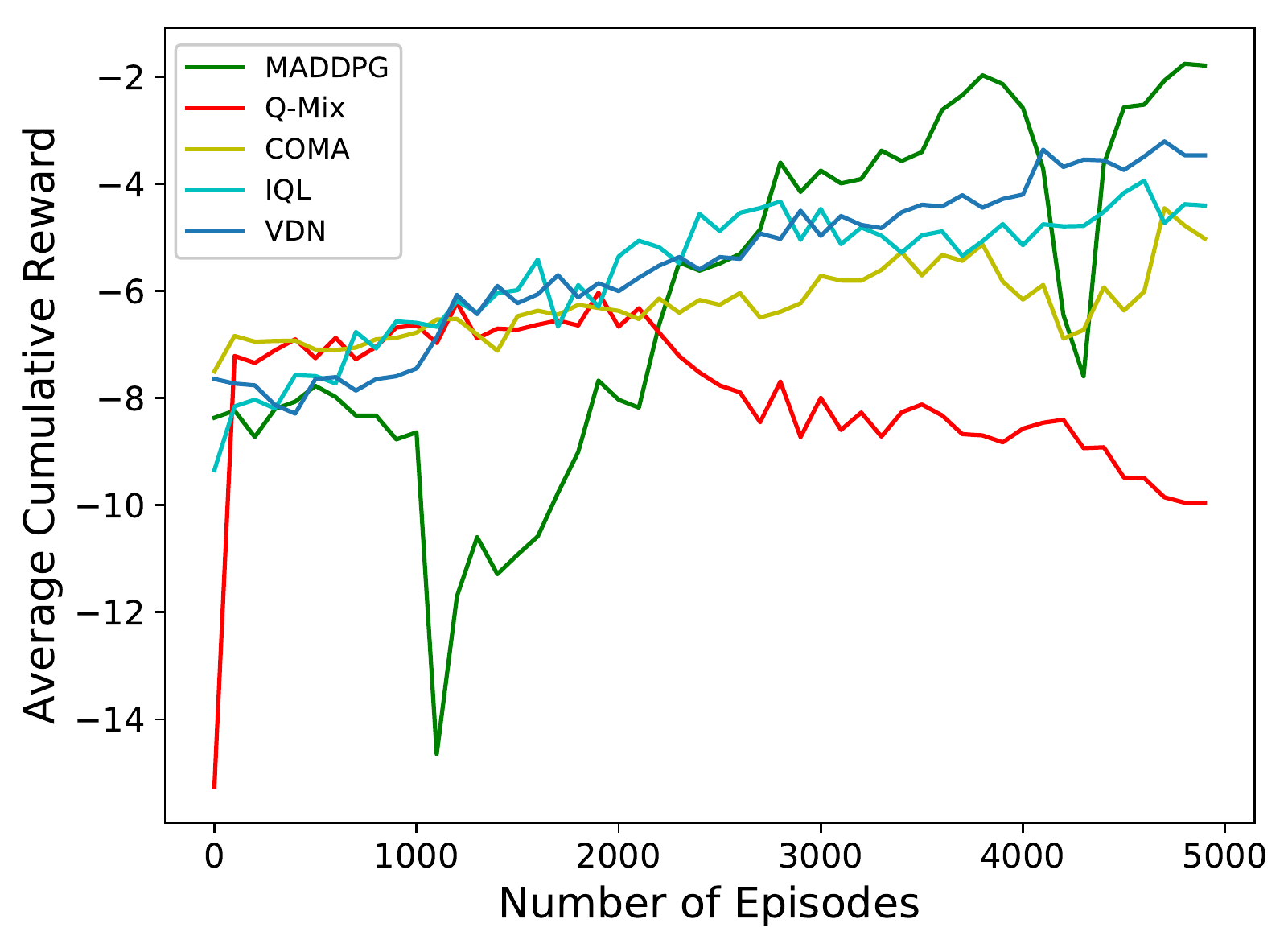}
        \caption{Red Carpet \label{subfig:rc_reward}}
    \end{subfigure}%
    \caption{Learning curve of the scenarios in terms of average cumulative reward for the bodyguards. Notice that start-of-the-art multi-agent reinforcement learning algorithms such as COMA and Q-MIX fail to even take off in most of the challenging scenarios. MADDPG was the only consistent algorithm that was able to learn in the environment.}
    \label{fig:average_reward}
\end{figure*}

\section{Experimental Setup}
\label{sec:experimental_setup}

\subsection{The VIP Protection Problem}
To demonstrate the effectiveness of our proposed algorithm, we simulated an urban security problem of VIP protection where a team of learning agents (bodyguards) are providing physical protection to a VIP from bystanders in a crowded space. This problem is briefly explored in~\cite{Sheikh-2018-GoalsRL,Sheikh-2018-AIT}.
\label{subsec:ModelingBodyguard}

We are considering a VIP moving in a crowd of bystanders $\mathcal{B}=\left\{ b_{1},b_{2},\ldots,b_{M}\right\}$ protected from assault by a team of bodyguards $R=\left\{ r_{1},r_{2},\ldots,r_{N}\right\}$. To be able to reason about this problem, we need to quantify the {\em threat} to the VIP at a given moment from the nearby bystanders--the aim of the bodyguards is to reduce this value.

Two agents $x$ and $y$ have a {\em line of sight} $LoS(x,y)\in\left[0,1\right]$ if $x$ can directly observe $y$ and with no obstacle between them. A bystander $b$ can only pose a threat to the VIP if it is closer than the safe distance $\mathrm{SafeDist}$. The {\em threat level} $\mathrm{TL}\left(\mathrm{VIP},b\right)$~\cite{Bhatia-2016-FLAIRS} is defined as the probability that a bystander $b$ can successfully assault the VIP, defined as a value exponentially decaying with distance:
\begin{equation}
\label{eq:ct}
\mathrm{TL}\left(\mathrm{VIP},b\right)=\exp^{-A\left(\mathrm{Dist}(\mathrm{VIP},b)\right)/B}
\end{equation}
\noindent where the VIP should be in line of sight of $b$ and  $\mathrm{Dist}(\mathrm{VIP},b)<\mathrm{SafeDist}$. $A$ and $B$ are positive constants that control the decay rate of the threat level.


The residual threat $RT\left(VIP, \mathcal{B}, R\right)$ is defined as the threat to the VIP at time $t$ from bystanders $\mathcal{B}$. Bodyguards can block the line of sight from the bystanders, thus the residual threat is always smaller than the threat level and depends on the position of the bodyguards with respect to the bystanders and the VIP. The {\em cumulative residual threat} to the VIP from bystanders $\mathcal{B}$ in the presence of bodyguards $R$ over the time period $[0,T]$ is defined as:
\begin{equation}
\label{eq:crt}
\mathrm{CRT}=\intop_{0}^{T}1-\prod_{i=1}^{k}\left(1-RT\left(VIP,b_{i}, R\right)\right)dt
\end{equation}
Our end goal is to minimize $\mathrm{CRT}$ through multi-agent reinforcement learning.

\subsection{Simulation and Scenarios}
\label{subsec:simulation_and_scenarios}
We designed four scenarios inspired from possible real world situations of VIP protection and implemented them as behaviors in the Multi-Agent Particle Environment(~\cref{fig:environments})~\cite{Mordatch-2017-ARXIV}.

In each scenario, the participants are the VIP, 4 bodyguards and 10 bystanders of one or more classes. The scenario description contains a number of {\em landmarks}, points on a 2D space that serve as a starting point and destinations for the goal-directed movement by the agents. For each scenario, the VIP (brown disk) starts from the starting point and moves towards the destination landmark (green disk). The VIP exhibits a simple path following behavior, augmented with a simple social skill metric: it is about to enter the personal space of a bystander, it will slow down or come to a halt.

The scenarios differ in the arrangement of the landmarks and the behavior of the different classes of bystanders.



\begin{enumerate}[label=\textbf{\Alph*}]
\item \textbf{Random Landmark:} In this scenario, 12 landmarks are placed randomly in the area. The starting point and destination for the VIP are randomly selected landmarks. The bystanders are performing random waypoint navigation: they pick a random landmark, move towards it, and when they reached it, they choose a new destination.  A set of fixed seeds were used for placement of landmarks and a different set of seeds were used for spawning bystanders in the environment.


\item \textbf{Shopping Mall:} In this scenario, 12 landmarks are placed in fixed position on the periphery of the area, representing shops in a market. The bystanders visit randomly selected shops and were spawned using a fixed set of random seeds.

\item \textbf{Street:} This scenario aims to model the movement on a crowded sidewalk. The bystanders are moving towards waypoints that are outside the current area. However, due to their proximity to each other, the position of the other bystanders influence their movement described by laws of particles motion~\cite{Vicsek-1995-PRL}.


\item \textbf{Pie-in-the-Face}: While the in other scenarios the bystanders treat the VIP as just another person, in this ``red carpet'' scenario the bystanders take an active interest in the VIP. We consider two distinct classes of bystanders with different behaviors. {\em Rule-abiding} bystanders stay behind a designated line observing as the VIP passes in front of them. {\em Unruly} bystanders break the limit imposed by the line and try to approach the VIP (presumably, to throw a pie in his/her face).

\end{enumerate}



\subsubsection*{Observation and Action Space}
\label{subsec:representations}
Following the model of Multi-Agent Particle Environment~\cite{Mordatch-2017-ARXIV}, the action space $\mathcal{A}_i$ of each bodyguard $i$ consists of 2D vector of forces applied on the bodyguard and to promote collaboration and cooperation~\cite{Mordatch-2017-ARXIV}, a \textbf{c} dimensional communication channel.

The observation of each bodyguard is the physical state of the nearest $m \subset M$ bystanders, all the $N$ bodyguards in the scenario and their verbal utterances such that $o_{i}=\left[x_{j,\ldots N+m}, c_{k,\ldots N}\right]\in \mathcal{O}_i$ where $x_{j}$ is the observation of the entity $j$ from the perspective of agent $i$ and $c_{k}$ is the verbal utterance of the agent $k$.

In this problem, we are assuming that all bodyguards have identical observation space and action space. Moreover, each scenario embedding $g$ is represented as a one hot vector.

\subsection{Reward Function}
\label{subsec:RewardFunctions}

Using the definitions of {\em threat level} and {\em cumulative residual threat} defined in~\cref{eq:ct} and~\cref{eq:crt} respectively, the reward function $r_{b}$ for bodyguard $i$
can be written as

\begin{equation}
r_{b} =-1+\prod_{i=1}^{k}\left(1-\mathit{RT}\left(VIP,b_{i}, R\right)\right)
\label{eq:ThreatOnly}
\end{equation}
To encourage the bodyguards to stay at a limited distance from the VIP and discourage them to attack the bystanders, a distance regularizer $\mathcal{D}$ is added to~\cref{eq:ThreatOnly} to form the final reward function.

\begin{equation}
\label{eq:final_reward_function}
\mathcal{D}\left(\mathit{VIP},x_{i}\right)=\begin{cases}
0 & m \leq\left\Vert x_{i}-\mathit{VIP}\right\Vert _{2}\leq d\\
-1 & \text{otherwise}\\
\\
\end{cases}
\end{equation}

\noindent where $m$ is the minimum distance the bodyguard has to maintain from VIP and $d$ is the $\mathrm{SafeDist}$ mentioned in~\cref{subsec:ModelingBodyguard}.  The final reward function is represented as

\begin{equation}
\begin{aligned}
r_{b}  =&\alpha\left(\displaystyle -1+\prod_{i=1}^{k}\left(1-\mathit{RT}\left(VIP,b_{i}, R\right)\right) \right) \\&
+\beta\left( \mathcal{D}\left(\mathit{VIP},x_{i}\right)\right)
\end{aligned}
\label{eq:Composite}
\end{equation}

Depending upon on the scenario $g$, different values of $\alpha$, and $\beta$ were chosen for the optimal performance. A different value of $\alpha$ and $\beta$ also fulfills the requirement of a different reward function to train a UVFA.

\section{Experiments and Results}
\label{sec:experiments}
In this section, we first evaluate the usefulness of multi-agent reinforcement learning
on the given problem by comparing it's results with an hand-engineered solution for the VIP problem. Then we demonstrate the inability of the state-of-the-art MARL algorithms to generalize over different scenarios. Finally we compare the results of scenario dependant policies with the results of universal policies. Our primary evaluation metric is {\em{Cumulative Residual Threat (CRT)}} defined in~\cref{eq:crt}.
\subsection{Multi-Agent Reinforcement Learning vs Quadrant Load Balancing}
\label{subsec:qlb}

In order to verify that multi-agent reinforcement learning solutions are better than explicitly programmed behavior of the bodyguards, we evaluate policies trained on individual scenarios with quadrant load balancing technique (QLB) introduced in~\cite{Bhatia-2016-FLAIRS}.

To identify the best MARL algorithm to compete with the hand-engineered solution, we trained five state-of-the-art MARL algorithms such as Q-Mix, VDN, IQL, COMA\footnote{https://github.com/oxwhirl/pymarl/} and MADDPG on our environment. It can be seen in~\cref{fig:average_reward} that MADDPG was the only algorithm that was successful in learning in our environment while other algorithms fail even to take off. Therefore,  we dropped the CRT graphs.

\begin{figure}
        \centering
        \includegraphics[height=1.75in, width=3.40in]{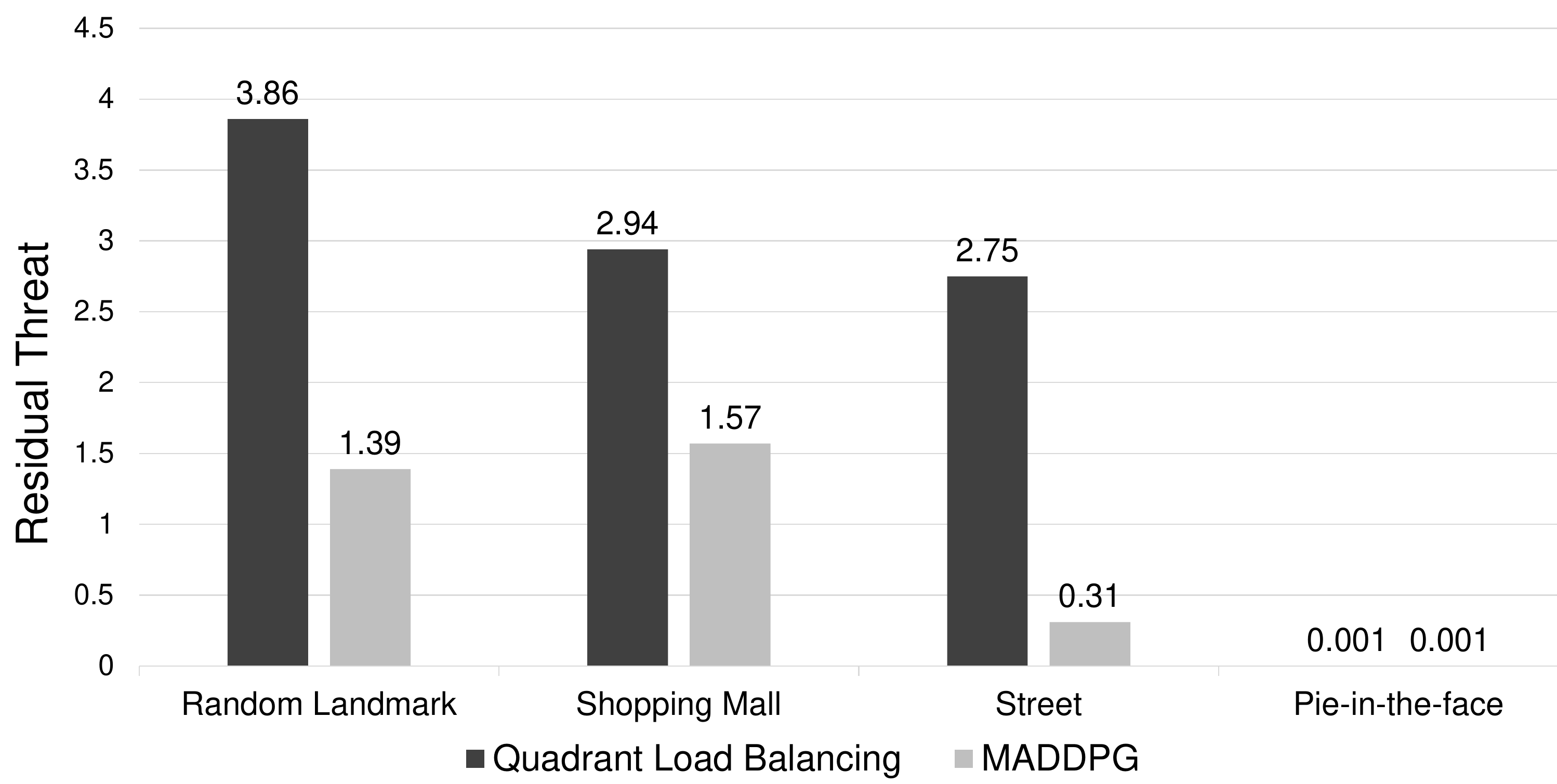}
\caption{Comparing the average cumulative residual threat values of MADDPG and QLB on four different scenarios.}
\label{fig:QLBRL}
\end{figure}

We then compared the results of MADDPG with quadrant load-balancing (QLB). From the results in~\cref{fig:QLBRL} we can see that the outcome are different depending on the characteristics of the scenario. For the Pie-in-the-face scenario, where most of the bystanders stay away behind the lines, both the RL-learned agent and the QLB model succeeded to essentially eliminate the threat. This was feasible in this specific setting, as there were four bodyguard agents for one ``unruly'' bystander. For the other scenarios, the average cumulative residual threat values are higher for both algorithms. However, for the Random Landmark and Shopping Mall scenarios the RL algorithm is able to reduce the threat to less than half, while in the case of the Street scenario, to less than one ninth of the QLB value.

Overall, these experiments demonstrate that the multi-agent reinforcement learning can learn behaviors that improve upon algorithms that were hand-crafted for this specific task.

\subsection{Universal Policies Vs Scenario-Dependant Policies}
\label{subsec:unified}
In order to verify the claim that MARL algorithms trained on specific scenario fail to generalize over different scenarios, we evaluate policies trained via MADDPG on specific scenario and test them on different scenarios. Policies on specific scenarios were trained using the same settings and configurations from experiments performed in~\cref{subsec:qlb}.

\begin{figure}[t]
        \centering
        \includegraphics[height=1.75in, width=3.40in]{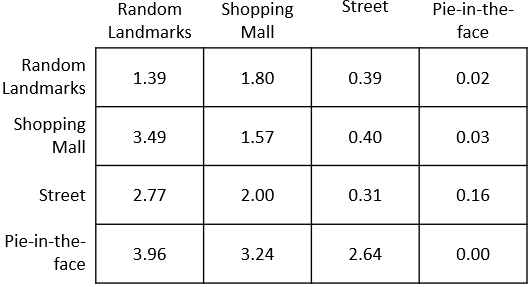}
\caption{A confusion matrix representing the average residual threat values of MADDPG policies trained on specific scenario when tested on different scenarios over 100 episodes.}
\label{tbl:cm}
\end{figure}

\begin{figure}
    \centering
    \includegraphics[height=1.9in, width=3.4in]{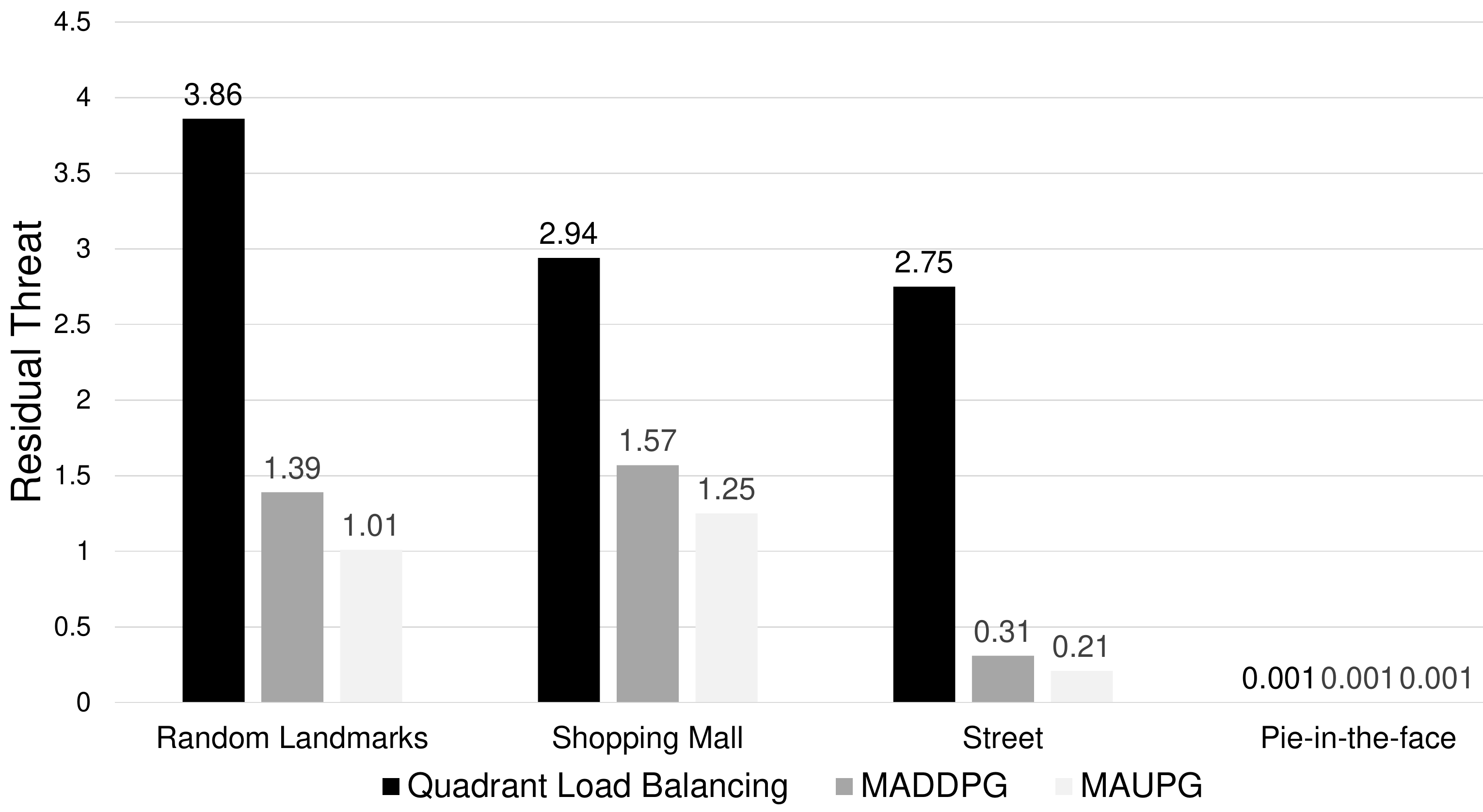}
    \caption{Comparing the average cumulative residual threat values for universal policy agents with MADDPG and QLB agents}
    \label{fig:combined}
\end{figure}

\noindent From the results shown in~\cref{tbl:cm} we can see that MADDPG policies trained on specific scenarios performed poorly when tested on different scenarios as compared to when tested on same scenario with different seeds.
In order to tackle the generalization problem, we train the agents using multi-agent universal policy gradient and compare its results with the results of scenario-dependant MADDPG policies.

From the results in~\cref{fig:combined} we can see that our proposed method performs better than policies trained on specific scenarios as well as quadrant-load balancing.
Overall, these experiments demonstrate that start-of-the-art MARL algorithms such as MADDPG fail to generalize a single task over multiple scenarios while our proposed solution MAUPG learn policies that allows a single task to be learnt across multiple scenarios and improve upon the start-of-the-art multi-agent reinforcement learning algorithm.

\section{Ablation Study}
\label{subsec:ablation}
The first natural question that can be asked here is that{ \em{why can't we just sample scenarios during training of a standard MADDPG?}} To answer this question and to see the effect of UVFA and hindsight replay used in MAUPG, we perform an ablation study in which we gradually add important building blocks to MADDPG to transform the solution into MAUPG.
First to answer the question, we trained a MADDPG and sampled different scenarios. Second we replaced the standard centralized critic with an UVFA and finally we added the hindsight replay step. All the training settings and hyperparameters were kept same across all the experiments.

\begin{figure}[H]
    \centering
        \includegraphics[height=1.9in, width=3.4in]{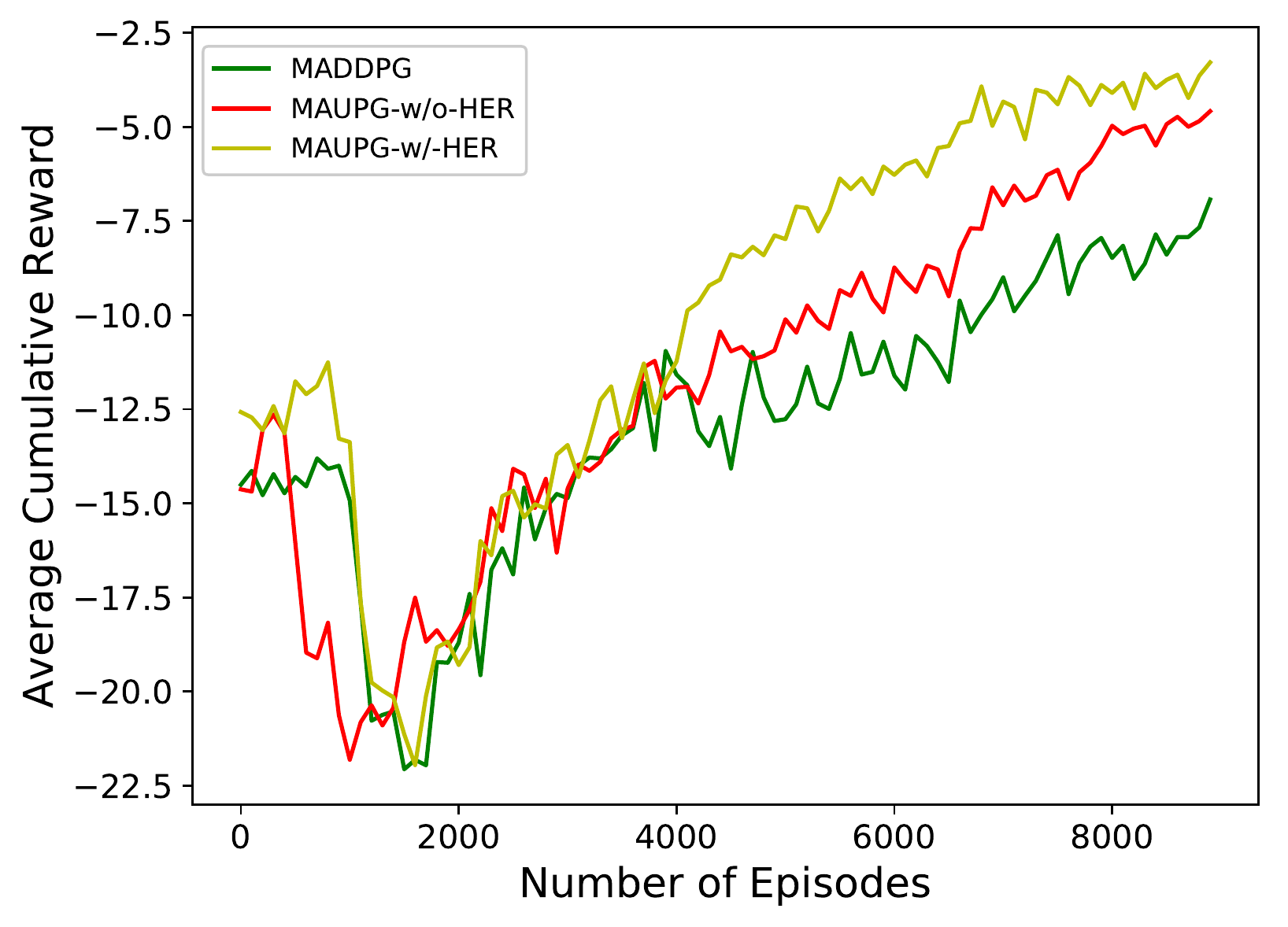}
        \caption{Learning curves of the ablated version of MAUPG. Notice the increase in performance in terms of average cumulative reward with an addition of UVFAs and hindsight replay step.  \label{fig:ablation_reward}}
\end{figure}

From ~\cref{fig:ablation_reward,fig:ablation_threat}, we can see that MADDPG does not learn half as good as MAUPG with or without the hindsight replay step. MAUPG learns better and faster than MAUPG without hindsight replay. This happens because MAUPG with hindsight replay benefits from replaying trajectories from one scenario in other scenarios(see lines~\ref{alg:hr_reward} and~\ref{alg:hr} of~\cref{alg:1}) thus providing more experience to learn efficiently.

\begin{figure}[H]
    \centering
     \label{fig:degenerate_threat}
        \includegraphics[height=1.9in, width=3.4in]{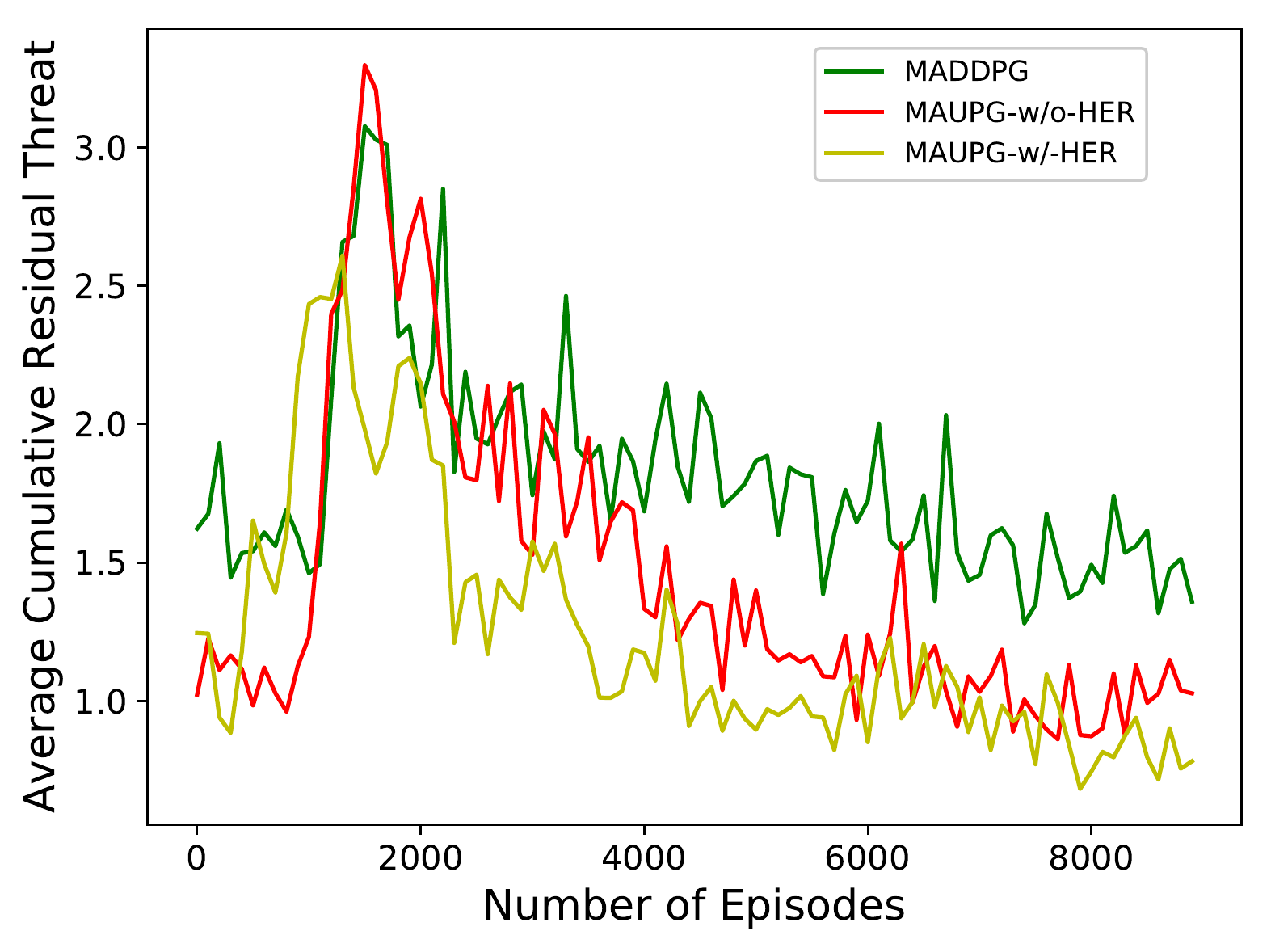}
        \caption{Learning curves of the ablated version of MAUPG. Notice the decline in average cumulative residual threat with an addition of UVFAs and hindsight replay step. \label{fig:ablation_threat}}
        \label{fig:ablation_threat}
\end{figure}

\section{Conclusion}
\label{sec:Conclusions}
In this paper, we highlighted the issue with MARL algorithms of failing to generalize a single task over multiple known scenarios.  
 To solve the generalization problem, we proposed {\em multi-agent universal policy gradient}, a universal value function approximator inspired policy gradient method that not only generalizes over state space but also over set of different scenarios. We also built a 2D challenging environment simulating an urban security problem that can be used as a benchmark for similar problems.  Experimental studies have demonstrated that our proposed method generalizes well on different scenario and performs better than MADDPG when trained on different scenarios individually.

\bibliography{ref}
\bibliographystyle{named}

\clearpage

\end{document}